\documentclass[twocolumn]{aastex631}

\usepackage{mathrsfs }
\usepackage{natbib}
\usepackage{amssymb}
\usepackage{apjfonts}
\usepackage{ulem} 
\usepackage{graphicx}
\begin{document}
\clearpage

\title{JWST/NIRSpec Balmer-line Measurements of Star Formation and Dust Attenuation at $\lowercase{z}\sim 3-6$}

\author[0000-0003-3509-4855]{Alice E. Shapley}\affiliation{Department of Physics \& Astronomy, University of California, Los Angeles, 430 Portola Plaza, Los Angeles, CA 90095, USA}
\email{aes@astro.ucla.edu}

\author[0000-0003-4792-9119]{Ryan L. Sanders}\altaffiliation{NHFP Hubble Fellow}\affiliation{Department of Physics and Astronomy, University of California, Davis, One Shields Ave, Davis, CA 95616, USA}

\author[0000-0001-9687-4973]{Naveen A. Reddy}\affiliation{Department of Physics \& Astronomy, University of California, Riverside, 900 University Avenue, Riverside, CA 92521, USA}

\author[0000-0001-8426-1141]{Michael W. Topping}\affiliation{Steward Observatory, University of Arizona, 933 N Cherry Avenue, Tucson, AZ 85721, USA}

\author[0000-0003-2680-005X]{Gabriel B. Brammer}\affiliation{Cosmic Dawn Center (DAWN), Denmark}\affiliation{Niels Bohr Institute, University of Copenhagen, Lyngbyvej 2, DK2100 Copenhagen \O, Denmark}

\shortauthors{Shapley et al.}

\shorttitle{Balmer lines at $z\sim 3-6$}

\begin{abstract}
We present an analysis of the star-formation rates (SFRs) and dust attenuation properties of star-forming galaxies at $2.7\leq z<6.5$ drawn from the Cosmic Evolution Early Release Science (CEERS) Survey. Our analysis is based on {\it JWST}/NIRSpec Micro-Shutter Assembly (MSA) $R\sim1000$ spectroscopic observations covering approximately $1-5$$\mu$m. Our primary rest-frame optical spectroscopic measurements are H$\alpha$/H$\beta$ Balmer decrements, which we use as an indicator of nebular dust attenuation. In turn, we use Balmer decrements to obtain dust-corrected H$\alpha$-based SFRs (i.e., SFR(H$\alpha$)). We construct the relationship between SFR(H$\alpha$) and stellar mass ($M_*$) in three bins of redshift ($2.7\leq z< 4.0$, $4.0\leq z< 5.0$, and $5.0\leq z<6.5$), which represents the first time the star-forming main sequence has been traced at these redshifts using direct spectroscopic measurements of Balmer emission as a proxy for SFR. In tracing the relationship between SFR(H$\alpha$) and $M_*$ back to such early times ($z>3$), it is essential to use a conversion factor between H$\alpha$ and SFR that accounts for the subsolar metallicity prevalent among distant galaxies. We also use measured Balmer decrements to investigate the relationship between dust attenuation and stellar mass out to $z\sim6$.  The lack of significant redshift evolution in attenuation at fixed stellar mass, previously confirmed using Balmer decrements out to $z\sim2.3$, appears to hold out to $z\sim 6.5$. Given the rapidly evolving gas, dust, and metal content of star-forming galaxies at fixed mass, this lack of significant evolution in attenuation provides an ongoing challenge to explain. 

\end{abstract}

\section{Introduction}
\label{sec:intro}

Hydrogen Balmer-line emission from H~II regions
has long been recognized as one of the most
robust probes of star formation and dust extinction in star-forming galaxies.
The Balmer decrement based on the H$\alpha$/H$\beta$ flux ratio can
be used to infer the amount of nebular attenuation, and, in turn, the dust-corrected, 
instantaneous star-formation rate (SFR) \citep[e.g.,][]{kennicutt1998}. The flux of a Balmer line, in combination with
the UV continuum flux density, can also be used to infer the efficiency
of ionizing photon production \citep[e.g.,][]{shivaei2018}, and search for evidence of 
bursty past star-formation histories \citep[e.g.,][]{dominguez2015,guo2016,emami2019,atek2022}. 

Vast samples of galaxies with multiple Balmer emission line measurements exist in the local universe,
from surveys such as the Sloan Digital Sky Survey \citep[SDSS;][]{abazajian2009}, and including
both integrated spectra and spatially-resolved emission-line maps \citep[e.g.,][]{belfiore2018,ellison2018}.
Large samples of Balmer decrements and dust-corrected H$\alpha$ SFRs (SFR(H$\alpha$)) 
were assembled for the first time at $z>1$ with the advent of
the {\it HST}/WFC3 IR grism \citep{dominguez2013,price2014,battisti2022} as well as multi-object near-IR spectrographs
on 8--10-meter class ground-based telescopes \citep{reddy2015}.
These measurements were used to trace the so-called ``main sequence" of galaxy formation
during the epoch of peak SFR density in the universe \citep{shivaei2015}, constrain
the nature of nebular dust attenuation and ISM geometry \citep{reddy2015,reddy2020,shivaei2020},
describe the spatially-resolved growth of galaxy disks \citep{nelson2016}, and investigate
the relationship between dust attenuation and stellar mass \citep{shapley2022}.

Until recently, it was impossible to perform
such fundamental measures of the star-forming galaxy population past
$z\sim 3$, because of both Earth's atmosphere and a lack of the required
instrumentation. Indeed, H$\alpha$ shifts past the red edge of the near-IR $K$ band  (2.4$\mu$m) beyond a redshift of $z=2.65$. The launch of {\it JWST} and the capabilities of its NIRSpec instrument
\citep{ferruit2022} have transformed the ability to detect both H$\alpha$
and H$\beta$, respectively, out to $z\sim 6.5$ and $z\sim 9.3$. Recent NIRSpec observations
from the Cosmic Evolution Early Release (CEERS) program \citep{finkelstein2022a,finkelstein2022b}
showcase this ability beautifully, for the first time enabling Balmer decrement measurements based on H$\alpha$
and H$\beta$ fluxes for a large sample of galaxies at $z\sim 3-6$. Here we report
on these Balmer decrements, as well as their implications for the star-formation rates (SFR(H$\alpha$))
and dust attenuation in typical star-forming galaxies extending from ``cosmic noon" back
into the reionization epoch.

In \S\ref{sec:obs}, we describe
our observations, data reduction, measurements, and sample. In \S\ref{sec:results}, we present results on the observed
relationships between SFR(H$\alpha$) and stellar mass, and Balmer decrement and stellar mass, 
measured for the first time at $z\sim 3-6$. 
In \S\ref{sec:discussion}, we consider the implications of these new measurements and consider
future directions.
Throughout, we adopt cosmological parameters of
$H_0=70\mbox{ km  s}^{-1}\mbox{ Mpc}^{-1}$, $\Omega_m=0.30$, and
$\Omega_{\Lambda}=0.7$, and a \citet{chabrier2003} IMF.

\section{Observations and Sample}
\label{sec:obs}

\subsection{The CEERS NIRSpec Program}
\label{sec:obs-ceers}
We use publicly available medium-resolution NIRSpec Micro-Shutter Assembly (MSA) data from the
CEERS program
\citep[Program ID:1345][; Arrabal Haro et al., in prep.]{finkelstein2022a, finkelstein2022b}. The CEERS NIRSpec
observations we analyzed consist of 6 pointings in the AEGIS field,
all of which utilized the grating/filter combination of G140M/F100LP,
G235M/F170LP, and G395M/F290LP, which provide a spectral resolution of $R\sim 1000$
over the wavelength range approximately $1-5\mu$m.
For each pointing, each grating/filter combination was observed
for a total of 3107 sec, broken down into three exposures of 14 groups,
and adopting the NRSIRS2 readout mode. A 3-point nod pattern was adopted for
each observation, and each MSA ``slit" consisted of 3 microshutters.
Each of the 6 pointings contained between 52 and 55 targets, for a total sample
of 321 slits and 318 distinct targets (3 galaxies were observed on two pointings).

\begin{figure*}[t!]
\centering
\includegraphics[width=0.95\linewidth]{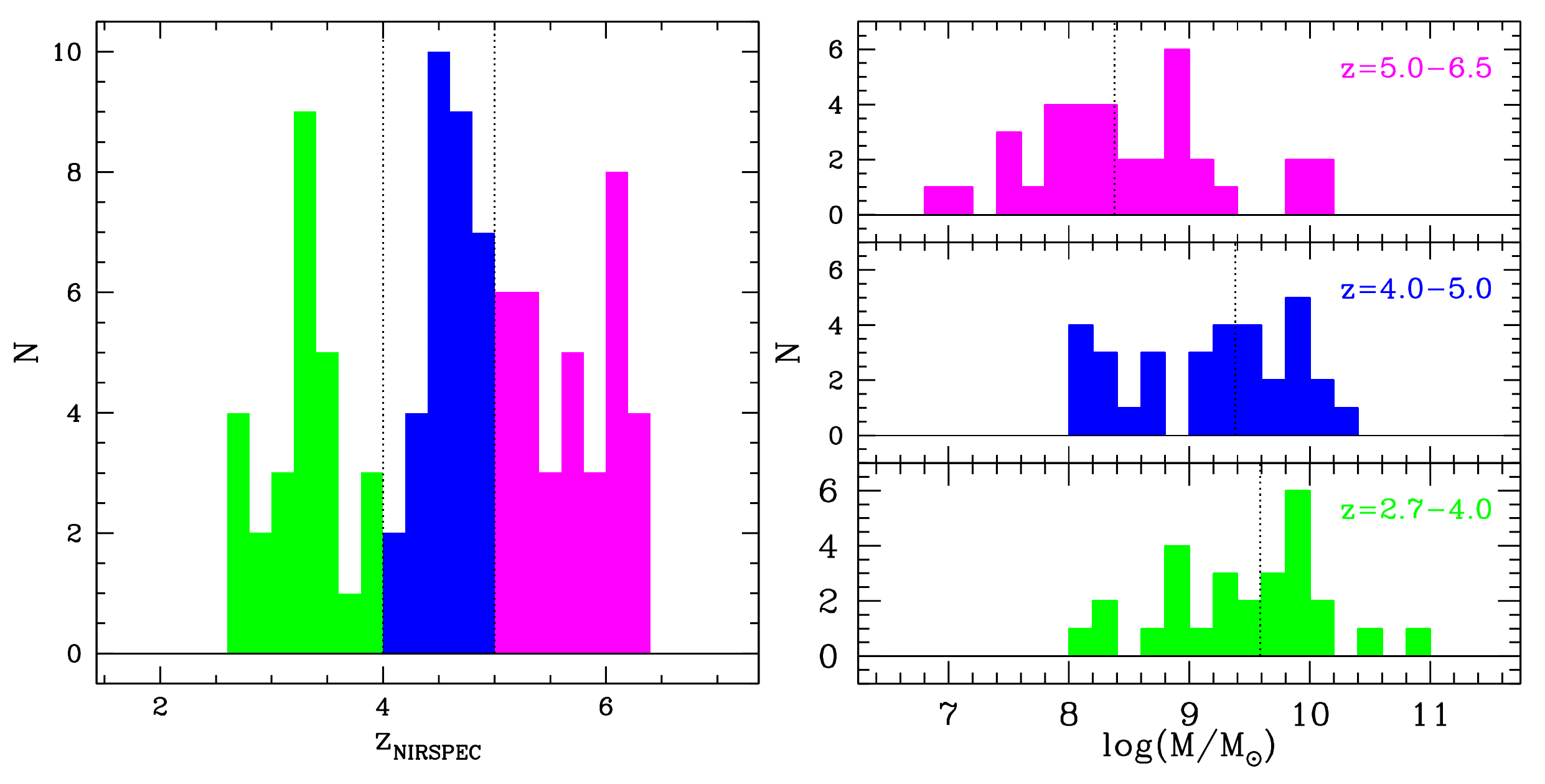}
\caption{{\bf Left:} Redshift distribution of all 113 CEERS galaxies at $2.7 \leq z \leq 6.5$, from which the sample of star-forming galaxies we analyze is drawn. The three redshift bins we delineate are indicated in green ($2.7 \leq z < 4.0$), blue ($4.0 \leq z < 5.0$), and magenta ($5.0 \leq z < 6.5$). {\bf Right:} Stellar mass distributions for the three redshift samples indicated in the left-hand panel, using the same color coding. For each redshift distribution, the median stellar mass is marked with a vertical dotted line. These median stellar mass values are $\log(M_*/M_{\odot})=$9.59, 9.38, and 8.38, respectively, for the $2.7 \leq z < 4.0$, $4.0 \leq z < 5.0$, and $5.0 \leq z < 6.5$ redshift samples.
}
\label{fig:sample}
\end{figure*}

\subsection{Data Reduction}
\label{sec:obs-redux}
We followed the same two-dimensional (2D) reduction procedures to reduce data for all three NIRSpec gratings.
We began by passing individual uncalibrated detector images through the JWST \texttt{calwebb\_detector1} pipeline \footnote{\url{https://jwst-pipeline.readthedocs.io/en/latest/index.html}}. In this step, we  masked all saturated pixels, subtracted the bias and dark current, and masked ``snowballs" and ``showers" associated with high-energy cosmic ray events. Images were then corrected for striping by estimating and subtracting the $1/f$ noise in each image. We then cut out the 2D spectrum for each MSA slit, and applied a flat-field correction, background subtraction using dithered exposures as the background, photometric calibration, and a wavelength solution based on the up-to-date calibration reference data system (CRDS) context (\texttt{jwst\_1027.pmap}). Each slitlet was rectified and interpolated onto a common wavelength grid based on its grating and filter combination. Finally, individual calibrated 2D spectra exposures were combined following the defined three-shutter dither pattern, while excluding pixels that had been previously masked. The 2D error spectra represent a combination of the variance from Poisson noise, read noise, flat-fielding, and variance between exposures, summed in quadrature. This stage of the reduction yielded 310 targets with 2D spectra covering all three gratings, reflecting a negligible sample of 8 initial targets that did not result in a viable 2D reduction.

One-dimensional (1D) science and error spectra were optimally extracted from the rectified 2D spectra \citep{horne1986}. The spatial profile in each grating was obtained by manually identifying wavelength ranges in the 2D spectrum containing high-S/N emission lines when present or detected continuum otherwise and summing the corresponding columns of the 2D spectrum.
For targets with detected lines or continuum in at least one grating, a blind extraction was applied to any remaining grating lacking such information. Out of 310 CEERS targets with the full set of 2D spectra, we extracted 1D spectra for 252 (i.e., 81\%). For the remaining 58 targets, in which we could identify neither lines nor continuum in any grating, we did not extract spectra, with either a manual or blind extraction technique.

As described in detail in \citet{reddy2023}, wavelength-dependent slit-loss corrections were estimated for each target based on its intrinsic morphology and position in the NIRSpec slit, as well as the wavelength-dependent {\it JWST} PSF. Intrinsic morphologies were estimated from {\it JWST}/NIRCam F115W imaging if available, or a S\'ersic fit to {\it HST}/F160W imaging if not. In the absence of NIRCam F115W imaging or a robust S\'ersic fit, a point source was assumed. 

As described in detail in \citet{sanders2023}, the final flux calibration was achieved by scaling 1D science spectra to match the photometric spectral energy distributions (SEDs). Slit-loss-corrected NIRSpec spectra were passed through the available photometric filter transmission for each target to produce synthetic photometric flux densities and errors.
The ratio of the image-based and synthetic flux densities was calculated for each filter in which both types of  measurements had S/N$>$5. If the number of filters meeting this requirement was $\geq 3$, 1D spectra and error spectra in all three
gratings were scaled by the median of the individual ratios to achieve the final flux calibration. For the 109 targets that did not meet this criterion, no scale factor was applied. For the remaining 143 targets, the median scale factor was 0.997 with a standard deviation of 0.23~dex.

\subsection{Measurements}
\label{sec:obs-measurements}
Redshifts and emission-line fluxes were measured from the 1D spectra for which we were able to robustly identify emission lines. Reported redshifts for 231 galaxies are based on the best-fit centroid from a single Gaussian fit to the line with the highest signal-to-noise ratio, usually [OIII]$\lambda$5007 (57\%) or H$\alpha$ (36\%). As described in more detail in \citet{sanders2023}, to estimate line fluxes, we used single Gaussian fits for widely-separated lines, adjacent lines such as [NII]$\lambda$6548, H$\alpha$, and [NII]$\lambda$6583 are fit simultaneously with multiple Gaussians, and closely spaced lines that are blended and unresolved at $R\sim1000$ are fit with a single Gaussian. The continuum model is taken to be the best-fit SED model (described below), where the only free parameter is an additive offset.
Using the best-fit SED model as the continuum has the advantage of self-consistently accounting for stellar absorption such that the measured hydrogen recombination line fluxes are robust. 

The same emission line was measured in two adjacent gratings for many targets. These overlapping measurements showed good agreement, with a median offset of 0.02~dex and an intrinsic scatter of 0.08~dex for all emission lines detected at $\geq3\sigma$ in both overlapping gratings, suggesting that the relative flux calibration between grating configurations is robust on average. In these cases of overlapping spectra, we adopted the inverse-variance weighted mean of the two available fluxes as our reported measurement.

We used existing multi-wavelength catalogs to derive best-fit SED models from which we infer stellar masses ($M_*$) and other stellar population parameters. Specifically, for the 99 CEERS NIRSpec targets with coverage, we used the
publicly available catalog constructed by G. Brammer\footnote{https://s3.amazonaws.com/grizli-v2/JwstMosaics/v4/index.html},
which includes 7 {\it HST} bands (F435W, F606W, F814W, F105W, F125W, F140W,
and F160W), and 7 {\it JWST}/NIRCam bands 
F115W, F150W, F200W, F277W, F356W, F410M, 
and F444W) from the initial CEERS NIRcam observations in June 2022.
For an additional 185 objects we used the spectral energy distributions
in the AEGIS field cataloged by the 3D-HST team \citep{momcheva2016,skelton2014},
which include ground-based and {\it HST} optical and near-IR photometry,
and measurements from {\it Spitzer}/IRAC at 3.6--8.0$\mu$m. There were
35 CEERS NIRSpec targets not covered by the Brammer HST+NIRCam catalog,
and lacking a robust multi-wavelength SED in the 3D-HST catalog. Restricted
to the sample of 231 galaxies with NIRSpec spectroscopic redshifts,
we found robust SED information for 210.
When restricted to the redshift range forming the basis of this analysis, i.e., $2.7 \leq z \leq 6.5$, we have robust SEDs for 94 galaxies (an additional 15 galaxies with measured spectroscopic redshifts in this range lack SED information).

For SED modeling, we used the FAST program \citep{kriek2009}, assuming
the stellar population synthesis models of \citet{conroy2009}, and a
\citet{chabrier2003} IMF. Following \citet{reddy2018a}, we adopted two combinations
of metallicity and extinction curves for SED modeling. These include 1.4 solar
metallicity ($Z_{\odot}=0.014$) coupled with the \citet{calzetti2000} attenuation curve (hereafter ``1.4 $Z_{\odot}+$Calzetti"), and 0.27 solar models coupled with
the SMC extinction curve of \citet{gordon2003} (hereafter ``0.27 $Z_{\odot}$+SMC"). We assumed
delayed-$\tau$ star-formation histories, where $SFR(t)\propto t \times exp(-t/\tau)$.
Here, $t$ is the time since the onset of star formation and $\tau$ is
the characteristic star-formation timescale. The adoption of
1.4 $Z_{\odot}+$Calzetti or 0.27 $Z_{\odot}$+SMC was determined
for each galaxy on the basis of its redshift and mass. Following
\citet{du2018} and guided by the evolving galaxy mass-metallicity relation
\citep[e.g.,][]{sanders2021}, at $z\leq 1.4$ we adopted 1.4 $Z_{\odot}+$Calzetti.
At $1.4 < z \leq 2.7$ ($2.7 < z \leq 3.4$), we adopted 1.4 $Z_{\odot}+$Calzetti for galaxies above
$\log(M_{*,1.4Z_{\odot}{\rm +Calzetti}}/M_{\odot})=10.45$ (10.66) and 0.27 $Z_{\odot}$+SMC for those at lower masses. At $z>3.4$, we adopted
0.27 $Z_{\odot}$+SMC models \citep{reddy2018a} for all stellar masses. The choice of dust laws described above applies to the full CEERS NIRSpec spectroscopic sample. When restricted to the redshift range probed in the current analysis ($2.7 \leq z \leq 6.5$) and given the range of stellar masses spanned by such targets, all but one galaxy is modeled using 0.27 $Z_{\odot}$+SMC.
We note that all relevant
photometric bands were corrected for the contributions from strong nebular
emission lines using the method described in \citet{sanders2021}, and Balmer emission-line fluxes were corrected
for the underlying stellar absorption implied by the best-fit stellar population
model.

Finally, SFR(H$\alpha$) was estimated from dust-corrected H$\alpha$ luminosities. \citet{reddy2020} showed that the Milky Way dust law of \citet{cardelli1989} provides a good match to the wavelength dependence of nebular attenuation in $z\sim 2.3$ star-forming galaxies. Accordingly, we used the measured H$\alpha$/H$\beta$ ratio, along with an assumption of the \citet{cardelli1989} dust extinction curve, to infer $E(B-V)_{\rm neb}$, the nebular extinction.  Then the dust-corrected H$\alpha$ luminosity was multiplied by a conversion factor depending on the metallicity of the best-fit SED model. Given the subsolar metallicity prevalent among high-redshift galaxies \citep[e.g.,][]{cullen2019,sanders2021}, it is essential to use a conversion factor between H$\alpha$ and SFR that accounts for this property. Following the analysis of \citet{reddy2018a}, for galaxies with 1.4 $Z_{\odot}+$Calzetti 
fits, we used a conversion factor of $10^{-41.37} (M_{\odot}{\rm yr}^{-1})/(\mbox{erg~s}^{-1})$, derived from $Z=0.02$ BPASS population synthesis models including the effects of stellar binaries and assuming an upper-mass IMF cut-off of  100~$M_{\odot}$. This calibration is almost identical to the one from \citet{hao2011} used in many other recent works for H$\alpha$ observations of $z\sim 2$ galaxies \citep{shivaei2015,sanders2021,shapley2022}. For galaxies with 0.27 $Z_{\odot}$+SMC fits, we used a conversion factor of $10^{-41.67} (M_{\odot}{\rm yr}^{-1})/(\mbox{erg~s}^{-1})$, derived from from $Z=0.001$ BPASS population synthesis models including the effects of stellar binaries and assuming an upper-mass IMF cut-off of  100~$M_{\odot}$ \citep{reddy2022}. The latter, lower conversion factor reflects the greater efficiency of ionizing photon production in lower-metallicity massive stars in binary systems.\footnote{The stellar metallicity associated with this  H$\alpha$ SFR conversion factor is lower than what is assumed for 0.27 $Z_{\odot}$+SMC broadband SED modeling, yet the conversion factor is not strongly metallicity-dependent in this low-metallicity regime.} 

\begin{figure}[t!]
\centering
\includegraphics[width=0.95\linewidth]{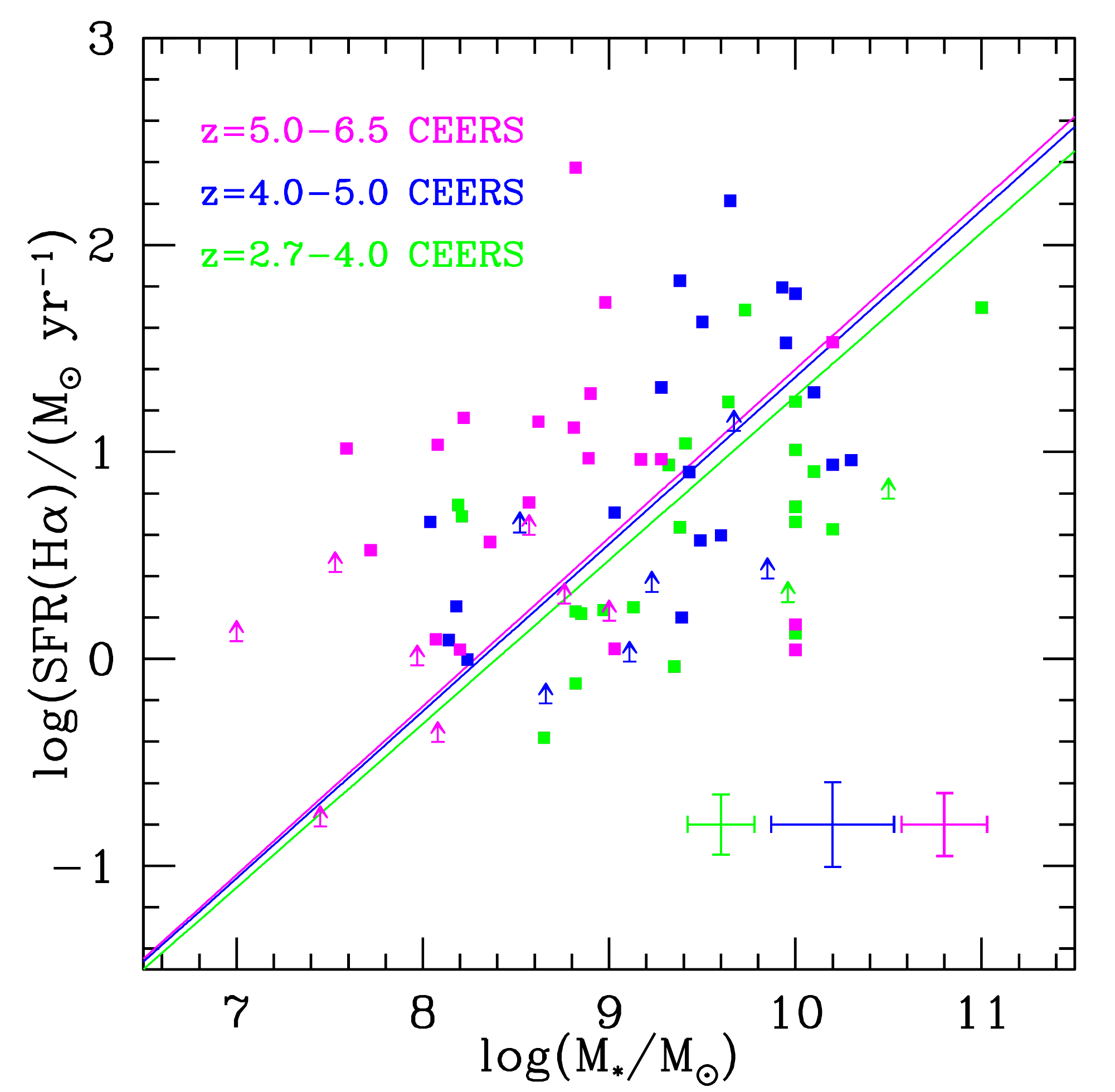}
\caption{SFR(H$\alpha$) vs. $M_*$. Green, blue, and magenta symbols are used, respectively, for the $2.7\leq z < 4.0$, $4.0 \leq z < 5.0$, and $5.0 \leq z < 6.5$ samples, and galaxies with H$\beta$ upper limits are indicated as SFR(H$\alpha$) {\it lower limits} (i.e., due to the lower limit on the Balmer decrement). The median error bar for each sample is shown in the lower-right corner of the plot in its designated color. Along with CEERS data points, we plot the best-fit relation from \citet{speagle2014} (their equation (28)), at the median redshift of each sample ($z=3.30, 4.60$ and $5.65$, respectively, for the $2.7\leq z < 4.0$, $4.0 \leq z < 5.0$, and $5.0 \leq z < 6.5$ samples), and offset by $-0.34$~dex in the $y$-axis to account for different assumptions regarding the conversion between observables and SFR.
}
\label{fig:sfrham}
\end{figure}

\begin{figure*}[t!]
\centering
\includegraphics[width=0.95\linewidth]{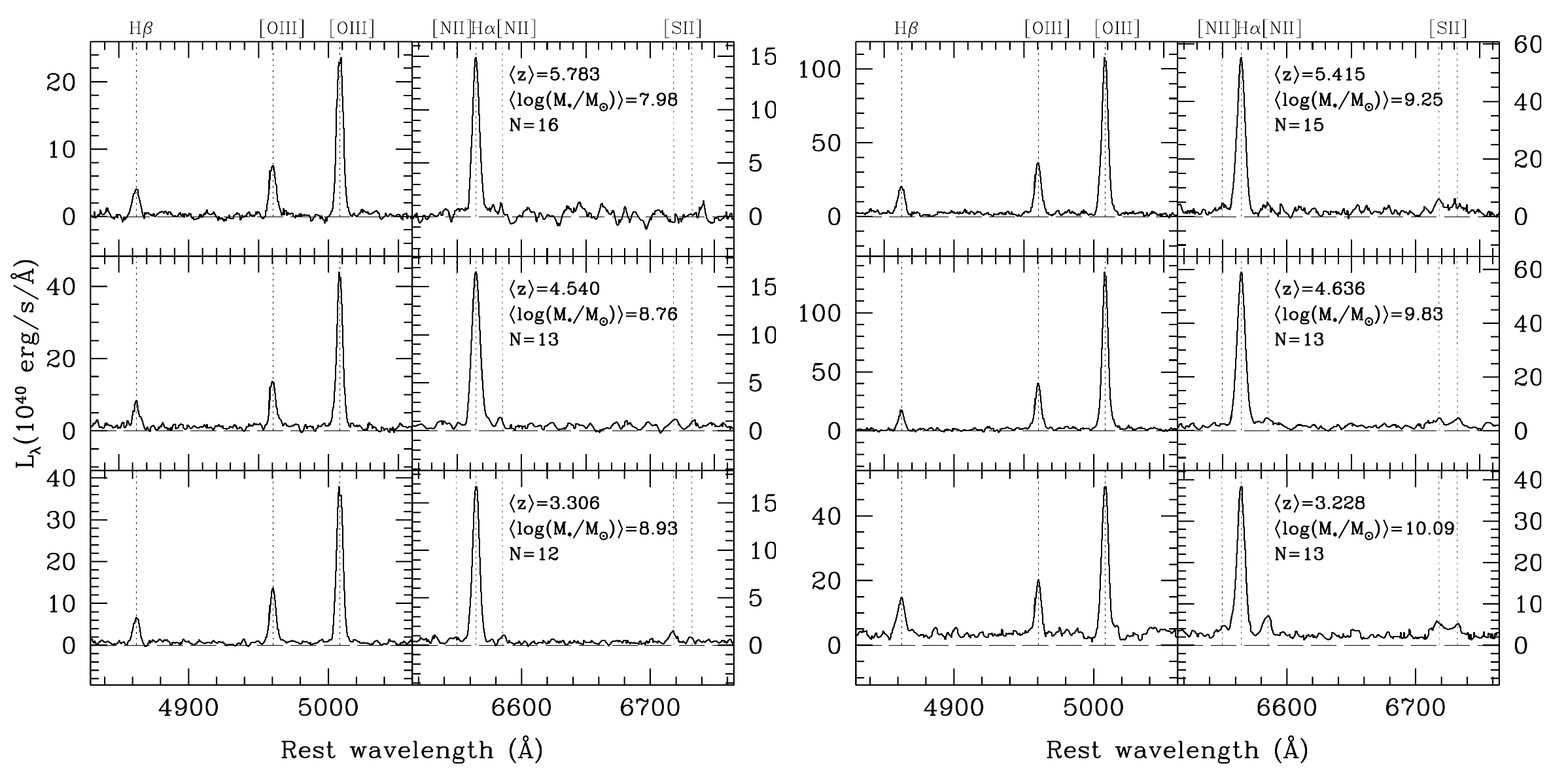}
\caption{Composite spectra for each of the three redshift bins, where, from bottom to top, we show spectra, respectively, for the $2.7\leq z<4.0$, $4.0\leq z < 5.0$, and $5.0\leq z < 6.5$ samples. 
In each row, the left set of panels represents the ``low-mass bin," while the right side indicates the ``high-mass bin," where each redshift sample is divided at the median stellar mass. 
Each composite spectrum is zoomed in on the regions covering H$\beta$ and [OIII]$\lambda\lambda 4959,5007$, as well as H$\alpha$, [NII]$\lambda\lambda 6548,6583$, and [SII]$\lambda\lambda 6717,6731$. These features are marked and labeled.
}
\label{fig:hahbplots}
\end{figure*}

\subsection{Sample}
\label{sec:obs-sample}
For the current analysis, we require a redshift measurement in the range $2.7 \leq z < 6.5$. The lower bound here represents the limit of ground-based measurements of H$\alpha$, i.e., the beginning of uncharted territory, while the upper bound represents the corresponding redshift limit imposed by the red cut-off of the G395M/F290LP setting. We also require a stellar mass estimate, wavelength coverage of both H$\alpha$ and H$\beta$, a $\geq 3\sigma$ detection of H$\alpha$, and finally a lack of indication of active galactic nucleus (AGN) activity. In the full sample of CEERS spectra, we identified 15 galaxies as candidate AGN on the basis of either an [NII]$\lambda6583$/H$\alpha$ ratio greater than 0.5 (10 galaxies), or else an H$\alpha$ profile consisting of both a narrow component and broad base (5 galaxies). 

Out of the 113 CEERS targets with redshifts measured at $2.7 \leq z < 6.5$ (Figure~\ref{fig:sample}, left), 109 show no rest-optical spectroscopic evidence for AGN activity, of which 94 have stellar mass estimates (Figure~\ref{fig:sample}, right). Of these galaxies, 77 have {\bf (1)} H$\alpha$ and H$\beta$ wavelength coverage and {\bf (2)} H$\alpha$ detections, and they comprise our primary sample. This primary sample is plotted in all subsequent figures that show datapoints for individual galaxies, and is representative of the larger sample of galaxies with measured redshifts in terms of redshift, stellar mass, and SFR as inferred from SED fitting. 
In order to search for evolution within the sample, we divide the primary sample into three redshift subsamples at $2.7 \leq z < 4$ (24 galaxies),  $4.0 \leq z < 5.0$ (25 galaxies), and $5.0 \leq z < 6.5$ (28 galaxies). In what follows, we plot individual galaxies that have both detections and upper limits for H$\beta$. The 62 galaxies with H$\beta$ detections (81\% of the primary sample) can be broken down into 22, 19, and 21 galaxies, respectively, at $2.7 \leq z < 4$,  $4.0 \leq z < 5.0$, and $5.0 \leq z < 6.5$.

\section{Results}
\label{sec:results}

\subsection{Star Formation}
\label{sec:results-sfr}

One of the key diagnostics of the evolution of the star-forming galaxy population across cosmic time is the so-called ``main sequence" \cite[e.g., ][]{noeske2007}. This correlation between SFR and $M_*$ is thought to reflect the gradual growth of galaxies, largely through smooth accretion and minor mergers. A galaxy's position with respect to the main sequence (within its scatter, significantly above, significantly below), provides a sense of its evolutionary state. 

In order to construct the SFR vs. $M_*$ relationship for CEERS galaxies targeted by NIRSpec, we took some care in translating dust-corrected H$\alpha$ luminosities. As described in Section~\ref{sec:obs-measurements}, across the entire CEERS NIRSpec spectroscopic sample, the adopted conversion factor is lower for lower-mass and higher-redshift galaxies, based on the observed trend towards lower metallicity at lower stellar mass and higher redshift. In fact, in our primary sample, all but one galaxy was modeled with a subsolar metallicity and SMC dust law. Accordingly, we used the low-metallicity SFR/$L_{\rm{H}\alpha}$ conversion factor for all but one galaxy as well. We note that the sample median SFR(H$\alpha$) estimated using this conversion factor shows excellent agreement (within 0.07~dex) with the median SFR derived from SED fitting (Section~\ref{sec:obs-measurements}), and the two sets of SFR measurements are significantly correlated \citep[see also][]{reddy2022}.

Figure~\ref{fig:sfrham} shows the relationship between  SFR(H$\alpha$) and $M_*$ among CEERS galaxies targeted by NIRSpec at $2.7 \leq z < 6.5$, color-coded by redshift range as in Figure~\ref{fig:sample}. We also plot the best-fit parameterized main sequence relation from \citet{speagle2014}, which expresses galaxy SFR as a function of both $M_*$ and $z$, or, equivalently, the age of the universe.  Relations from \citet{speagle2014} are plotted at the median redshift of each of the three subsamples ($z=3.3$, 4.6, and 5.65). Notably, we also shift the \citet{speagle2014} relations by $-0.34$~dex in SFR(H$\alpha$), since they are effectively tied to the \citet{hao2011} SFR conversion factor for H$\alpha$. 

Both the $2.7 \leq z < 4.0$ and $4.0 \leq z < 5.0$ samples scatter symmetrically around the (shifted) main sequence fits from \citet{speagle2014}, suggesting that these samples are representative of star-forming galaxies over the stellar mass range $8.0 \leq \log(M_*/M_{\odot}) \leq 10.0$. The two lower-redshift subsamples also show no significant offset with respect to each other in terms of typical SFR(H$\alpha$) at fixed $M_*$, consistent with the lack of strong redshift dependence in the \citet{speagle2014} over this redshift range. The $5.0\leq z < 6.5$ sample, however, is offset  towards higher SFR(H$\alpha$) relative to the \citet{speagle2014} parametrization, which, itself, represents an extrapolation out to such high redshifts. Regardless of the parametrized version of the main sequence, the highest-redshift subsample is characterized by a higher average SFR(H$\alpha$) at fixed stellar mass than the two lower-redhift subsamples within the stellar-mass range of overlap ($8.0 \leq \log(M_*/M_{\odot} \leq 9.0$). More representative samples will be required to determine if this offset is reflective of the underlying evolving galaxy population, or rather a selection effect.

\begin{figure*}[t!]
\centering
\includegraphics[width=0.95\linewidth]{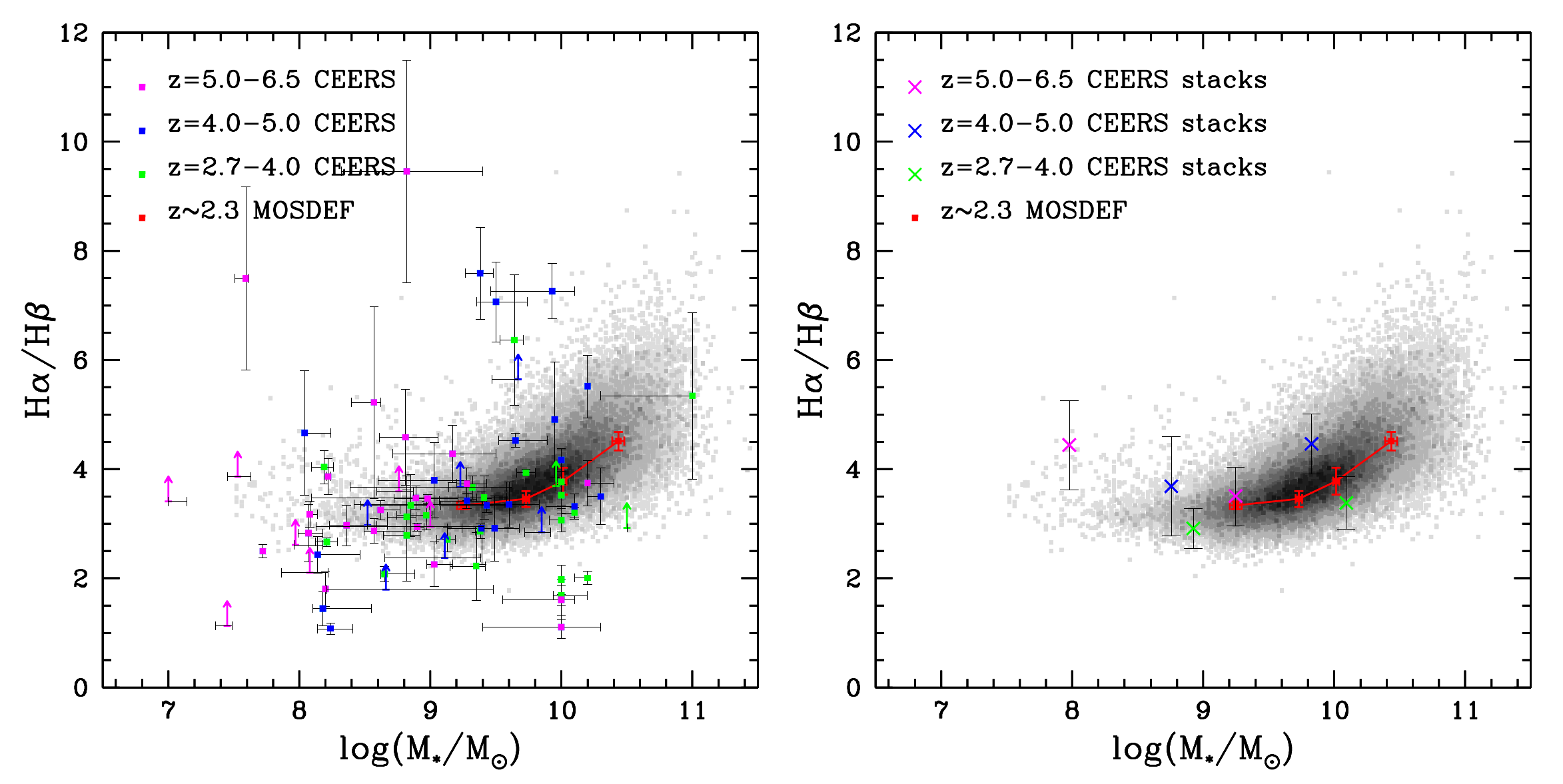}
\caption{Attenuation vs. $M_*$ based on the Balmer line ratio, H$\alpha$/H$\beta$. In each panel, the background grayscale histogram corresponds to the distribution of local SDSS galaxies. The running median H$\alpha$/H$\beta$ ratio for $z\sim 2.3$ star-forming galaxies in the MOSDEF survey is shown in red \citep{shapley2022}. In the left panel, we show individual CEERS galaxies color-coded by redshift as in previous plots. On the right, plotted H$\alpha$/H$\beta$ ratios are measured from composite spectra in two bins of stellar mass for each redshift range, as shown in Figure~\ref{fig:hahbplots}.}
\label{fig:hahblm}
\end{figure*}

\subsection{Dust Attenuation}
\label{sec:results-atten}

It has been shown that the strong connection between measures of dust attenuation and $M_*$ does not significantly evolve between $z\sim 0$ and $z\sim 2$. Here dust attenuation has been estimated with several different tracers, including the ratio of far-IR to UV SFRs or luminosities, also known as ``IRX" \citep[e.g.,][]{meurer1999,bouwens2016};
the magnitude of far-UV (i.e., 1600\AA) attenuation, or $A_{1600}$ \citep[e.g.,][]{mclure2018};
the fraction of star formation that is obscured, $f_{\rm obscured}$ \citep{whitaker2017},
and the nebular attenuation based on the Balmer decrement \citep[i.e., H$\alpha$/H$\beta$ ratio;][]{kashino2013,dominguez2013,price2014}.
There is less consensus regarding the form of the attenuation vs. $M_*$ relation at $z>3$,
with some evidence that it may evolve towards lower attenuation at fixed $M_*$ \citep[e.g.,][]{fudamoto2020}.

\citet{shapley2022} presented a large sample of Balmer decrements at $z\sim 2.3$, demonstrating that the relationship between Balmer decrement and $M_*$ showed no significant evolution up to the current epoch. We extend this work for the first time out to $z\sim 6.5$, using measured Balmer decrements for the CEERS NIRSpec sample. In addition to individual Balmer decrement measurements, we used stacked composite spectra to estimate average quantities in two bins of stellar mass for each of the three redshift bins. 
Composites were constructed by shifting
the spectrum of each individual galaxy in a bin into the rest frame, converting from flux density to luminosity density, sampling each individual spectrum on a common wavelength grid, scaling each spectrum to match the average H$\alpha$ luminosity in the bin, and then averaging luminosity densities at each wavelength point using 3$\sigma$ clipping. These composite spectra, zoomed in to the regions surrounding H$\beta$ and H$\alpha$, are shown in Figure~\ref{fig:hahbplots}. Each row features the results for one of the redshift subsamples, while the left-hand (right-hand) set of plots represents the lower-mass (higher-mass) half of each sample.\footnote{The sample for stacking ($N=82$) is slightly larger than for individual measurements, as there was no explicit requirement of H$\beta$ coverage in the stacks.}

The left-hand panel of Figure~\ref{fig:hahblm} shows the relationship between Balmer decrement and $M_*$ for both $z\sim 0$ star-forming galaxies in SDSS, and $z\sim 2.3$ galaxies drawn from the MOSDEF survey \citep{shapley2022}. We overplot individual CEERS NIRSpec measurements at  $2.7 \leq z < 6.5$, color-coded by redshift subsample. The individual $z\geq 3$ measurements are noisy, but there is no obvious evolution between $z\sim 2.3$ and $z\sim 6.5$. With the assumption of the \citet{cardelli1989} dust attenuation law, we can translate H$\alpha$/H$\beta$ to  $E(B-V)_{\rm neb}$, and find that $E(B-V)_{\rm neb}$ ranges from 0 to 1.19 in our $2.7 \leq z < 6.5$ sample, with a median of 0.18. We note that a small number of galaxies in the CEERS sample scatter to either surprisingly high values of H$\alpha$/H$\beta$, or else values significantly less than the dust-free minimum value of 2.79.\footnote{We adopt an intrinsic ratio of 2.79 for H$\alpha$/H$\beta$, which is based on assuming an electron temperature of $T_{e}=15000$~K. This temperature is typical of high-redshift, subsolar-metallicity, star-forming galaxies \citep{sanders2020,sanders2023,reddy2022,curti2023}.} We attribute {\it the majority of} these outliers to remaining systematics in the NIRSpec grating-to-grating flux calibration (i.e., when H$\alpha$ and H$\beta$ are measured in different gratings), which lacks bias on average, but also has scatter.

In the right-hand panel of Figure~\ref{fig:hahblm}, we replace individual CEERS measurements with those taken from composite spectra. In these higher-S/N measurements, the two lower-redshift samples show preliminary evidence for higher H$\alpha$/H$\beta$ ratio at higher stellar mass, yet the error bars are still too large to discern a significant trend. Furthermore, at lower redshifts, only a very shallow trend between H$\alpha/$H$\beta$ and stellar mass is observed over the stellar mass range probed by the $z>3$ sample. The main result from these preliminary measurements of dust attenuation and stellar mass in CEERS is that the $z>3$ measurements scatter around those at $z\sim 0-2.3$, with no obvious offset overall. We do note that the lower-mass bin at $5.0\leq z< 6.5$ is offset towards higher H$\alpha$/H$\beta$ relative to the SDSS distribution (there are no $z\sim 2.3$ measurements at  $\log(M_*/M_{\odot})\sim 8.0$) but the error bar for this 16-galaxy stack is large enough that its vertical offset relative to SDSS is not significant, and the lower- and higher-mass bins at this redshift are statistically consistent with a flat trend.

\section{Discussion}
\label{sec:discussion}

A crucial component of our measurement of the SFR(H$\alpha$) vs. $M_*$ relation at $2.7 \leq z < 6.5$ is the adoption of an appropriate conversion factor between dust-corrected H$\alpha$ luminosity and SFR, characterized by the correct metallicity and treatment of the effects of stellar binaries. There are multiple lines of evidence that the vast majority of galaxies in our sample have significantly subsolar metallicities. Their stellar masses alone suggest subsolar metallicity, given what is known about the evolution of the galaxy mass-metallicity relation at lower redshifts \citep[e.g.,][]{sanders2021}. For example, given the median stellar mass of our primary $2.7 \leq z < 6.5$ sample (i.e., $\log(M_*/M_{\odot})=$9.11), the corresponding oxygen abundance in the $z\sim 3.3$ mass-metallicity relationship of \citet{sanders2021} is $12+\log({\rm O/H})=8.15$ (i.e., 0.28 solar). Assuming that the mass-metallicity relationship evolves towards lower metallicity at fixed mass as redshift increases \citep[e.g.,][]{torrey2019}, we regard this metallicity as an upper limit for the typical value characterizing our primary sample. More directly, as discussed in \citet{shapley2023}, the composite spectra shown in Figure~\ref{fig:hahbplots} indicate [NII]/H$\alpha\leq 0.1$ for all subsamples, another sign of low metallicity \citep{pp2004}. Finally, models of the rest-UV stellar continuum of $z\sim 2-3$ star-forming galaxies suggests significantly subsolar stellar metallicities \citep{steidel2016,cullen2019,topping2020a,topping2020b,reddy2022}. Star-forming galaxies such as those in the CEERS NIRSpec sample, covering the same or lower stellar-mass range but at {\it higher} redshift, should be even {\it less} enriched. As highlighted by \citet{reddy2018b} and \citet{theios2019}, the subsolar conversion factor between H$\alpha$ luminosity and SFR adopted here is a factor of $\sim 2.5$ lower than the canonical conversion used for lower-redshift studies in the literature \citep[e.g.][]{hao2011}. Previously, \citet{caputi2017} estimated the SFR(H$\alpha$) vs. $M_*$ relation at $z\sim 4-5$, based on a large sample of star-forming galaxies with photometric redshifts and H$\alpha$ line fluxes inferred indirectly from {\it Spitzer}/IRAC 3.6$\mu$m photometric excesses relative to best-fit SED models. \citet{caputi2017} used a solar-metallicity conversion factor for SFR(H$\alpha$), resulting in a higher overall normalization of the SFR(H$\alpha$) vs. $M_*$ relation, and also found an apparent bimodality in the SFRs of galaxies with strong H$\alpha$ emission. We recover no such bimodality in the distribution of SFR(H$\alpha$) values, based on direct spectroscopic measurements of Balmer lines.

The CEERS NIRSpec sample provides tantalizing evidence that the relationship between Balmer decrement and stellar mass remains constant out to $z\sim 6.5$. This measure of dust attenuation depends on both the dust mass and the way in which it is distributed (i.e., the effective dust-mass surface density), so the lack of evolution in attenuation at fixed stellar mass suggests a constant ratio of dust-mass surface density to stellar mass \citep{shapley2022}. At the same time, other studies have found evidence for a lower fraction of obscured star-formation (``IRX") at fixed mass at $z>4$ \citep[e.g.][]{fudamoto2020}, based on far-IR and UV continuum estimate of dust attenuation. Different redshift evolution in the IRX vs. $M_*$ and H$\alpha$/H$\beta$ vs. $M_*$ relations could arise if the spatial distribution of dust relative to massive stars and H~II regions evolves \citep{reddy2015}, based on the fact that IRX probes stellar continuum attenuation while H$\alpha$/H$\beta$ traces nebular attenuation in H~II regions. However, the results thus far on the attenuation vs. mass relation at the highest redshifts -- both our Balmer decrement analysis and the studies based on IRX -- use small samples of galaxies, and require confirmation with an order of magnitude larger sample numbers. 

We have entered an era in which spectroscopic Balmer-line measurements at $z>3$ are routine and can be obtained in modest exposure times on {\it JWST}. The CEERS NIRSpec dataset analyzed here demonstrates the great potential of {\it JWST} for obtaining fundamental probes of the star-forming galaxy population into the reionization epoch based on Balmer-line measurements. In addition to tracing star formation, galaxy growth, and dust attenuation, as we do here, the ratio of H$\alpha$ to UV continuum luminosity can be used to infer the efficiency of ionizing photon production, $\xi_{{\rm ion}}$ \citep{shivaei2018} -- crucial for quantifying the role of star-forming galaxies in cosmic reionization -- as well as evidence for bursty star-formation histories \citep{emami2019}. To fully utilize Balmer line ratios at $z>3$ and robustly infer dust-corrected SFRs and other key galaxy properties, we must limit wavelength-dependent systematic uncertainties in NIRSpec flux calibration. We look forward to realizing the full potential of {\it JWST} and NIRSpec with not only larger and representative galaxy samples, but also samples with complete NIRCam photometric coverage, selected from early public {\it JWST} imaging datasets.

\section*{Acknowledgements}
We acknowledge the entire CEERS team for their effort to design and execute this Early Release Science
observational program, especially the work to design the MSA observations.
This work is based on observations made with the NASA/
ESA/CSA James Webb Space Telescope. The data were
obtained from the Mikulski Archive for Space Telescopes at
the Space Telescope Science Institute, which is operated by the
Association of Universities for Research in Astronomy, Inc.,
under NASA contract NAS5-03127 for JWST. The specific observations analyzed can be accessed via \dataset[DOI: 10.17909/z7p0-8481]{https://archive.stsci.edu/doi/resolve/resolve.html?doi=10.17909/z7p0-8481}
We also acknowledge support from NASA grant JWST-GO-01914. Support for this work was also provided through the NASA Hubble Fellowship
grant \#HST-HF2-51469.001-A awarded by the Space Telescope
Science Institute, which is operated by the Association of Universities
for Research in Astronomy, Incorporated, under NASA contract NAS5-26555.


\end{document}